\documentclass[twocolumn,aps]{revtex4-1}
\usepackage[latin9]{inputenc}
\usepackage{amsmath}
\usepackage{amssymb}
\usepackage{graphicx}

\makeatletter

\@ifundefined{textcolor}{}{%
\definecolor{BLACK}{gray}{0}
\definecolor{WHITE}{gray}{1}
\definecolor{RED}{rgb}{1,0,0}
\definecolor{GREEN}{rgb}{0,1,0}
\definecolor{BLUE}{rgb}{0,0,1}
\definecolor{CYAN}{cmyk}{1,0,0,0}
\definecolor{MAGENTA}{cmyk}{0,1,0,0}
\definecolor{YELLOW}{cmyk}{0,0,1,0}
}

\usepackage{upgreek}

\@ifundefined{textcolor}{}{%
\definecolor{BLACK}{gray}{0}
\definecolor{WHITE}{gray}{1}
\definecolor{RED}{rgb}{1,0,0}
\definecolor{GREEN}{rgb}{0,1,0}
\definecolor{BLUE}{rgb}{0,0,1}
\definecolor{CYAN}{cmyk}{1,0,0,0}
\definecolor{MAGENTA}{cmyk}{0,1,0,0}
\definecolor{YELLOW}{cmyk}{0,0,1,0}
}

\usepackage{color}

\def\NOT(#1,#2){\OneQubitGate(#1,#2){$X$}}

\@ifundefined{showcaptionsetup}{}{%
\PassOptionsToPackage{caption=false}{subfig}}

\makeatother

\begin{document}
\title{Magnetic field-assisted spectral decomposition and imaging of charge
states of NV centers in diamond}
\author{T. Chakraborty$^{1,2}$, R. Bhattacharya$^{3}$, V. S. Anjusha$^{3,4}$,
M. Nesladek$^{5,6}$, D. Suter$^{1,3}$ and T. S. Mahesh$^{3}$}
\affiliation{$^{1}$Fakult{ä}t Physik, Technische Universit{ä}t Dortmund, D-44221
Dortmund, Germany}
\affiliation{$^{2}$QuTech and Kavli Institute of Nanoscience, Delft University
of Technology, 2600 GA Delft, The Netherlands}
\affiliation{$^{3}$Department of Physics and NMR Research Center, Indian Institute
of Science Education and Research, Pune 411008, India}
\affiliation{$^{4}$Institut for Quantum Optics, Ulm University, Albert-Einstein-Allee
11, Ulm 89081, Germany}
\affiliation{$^{5}$Institute for Materials Research (IMO), Hasselt University,
B-3590 Diepenbeek, Belgium}
\affiliation{$^{6}$IMOMEC Division, IMEC, B-3590 Diepenbeek, Belgium}
\begin{abstract}
With the advent of quantum technology, nitrogen vacancy (NV) centers
in diamond turn out to be a frontier which provides an efficient platform
for quantum computation, communication and sensing applications. Due
to the coupled spin-charge dynamics of the NV system, knowledge about
NV charge state dynamics can help to formulate efficient spin control
sequences strategically. Through this paper we report two spectroscopy-based
deconvolution methods to create charge state mapping images of ensembles
of NV centers in diamond. First, relying on the fact that an off-axis
external magnetic field mixes the electronic spins and selectively
modifies the photoluminescence (PL) of NV$^{-}$, we perform decomposition
of the optical spectrum for an ensemble of NVs and extract the spectra
for NV$^{-}$ and NV$^{0}$ states. Next, we introduce an optical-filter
based decomposition protocol and perform PL imaging for NV$^{-}$
and NV$^{0}$. Earlier obtained spectra for NV$^{-}$ and NV$^{0}$
states are used to calculate their transmissivities through a long
pass optical filter. These results help us to determine the spatial
distribution of the NV charge states in a diamond sample.
\end{abstract}
\maketitle

\section{Introduction}

By the virtue of having remarkable quantum properties at room temperature
and spin dependent optical response, the nitrogen vacancy NV center
in diamond provides an efficient platform to implement protocols of
quantum technology \citep{Jelezko_review,Doherty:2013uq}. Long coherence
time \citep{balasubramanian2009ultralong}, photostable single photon
emitting capability \citep{kurtsiefer2000stable}, the possibility
of addressing and manipulating the spins via optical and microwave
excitation \citep{Childress_Science,Dutta_Science,chakraborty2017polarizing},
the possibility to readout the spin states by different methods \citep{hopper2018spin}
and efficient integrability into photonic structures \citep{sipahigil2016integrated}
have led to applications of NV centers in several aspects of quantum
technology. Notably, NV centers have exhibited promising applications
in quantum information processing \citep{Bernien_PRL,Neumann_Science},
magnetometry \citep{rondin2014magnetometry,wolf2015subpicotesla},
bio-sensing \citep{schirhagl2014nitrogen,haziza2017fluorescent},
thermometry \citep{wang2015high,neumann2013high} and so on. The negatively
charged state of the NV center (NV) is widely investigated allowing
optical and microwave excitation controlled preparation, manipulations
and read out of its spin states in an efficient way, whereas, qubits
associated with the neutral charge state (NV$^{0}$) and their quantum
control have not been well explored experimentally, although theoretical
proposals have been put forward\citep{felton2008electron,gali2009theory}.

In magnetometry applications, ensembles of NV$^{-}$ centers being
operated at ambient conditions have shown excellent efficiency in
field-imaging with high spatial resolution \citep{schirhagl2014nitrogen,casola2018probing}
and sensitivity up to 1 $\mathrm{pT/\sqrt{Hz}}$ \citep{wolf2015subpicotesla,barry2020sensitivity}.
An ensemble of $N$ NV centers can generate $N$ times the number
of photons per unit time compared to a single center and therefore
increase the sensitivity by a factor of $\sqrt{N}$ \citep{rondin2014magnetometry}.
Moreover, an ensemble, of NVs with four different crystal orientations,
allows performing direction-sensitive magnetic field sensing \cite{yahata2019demonstration}.
Although the NV$^{-}$ is the target state for sensing applications,
the sensitivity is reduced by the presence of neutral charge state
as NV$^{0}$centers add a spin-independent PL background to the desired
signal which affects the field detection sensitivity \citep{craik2020microwave}.
The NV photophysics results in an inter-conversion dynamics between
these two charge states \citep{aslam2013photo}. Thus NV$^{-}$ spin
properties and hence the sensing efficiency is affected by the coupled
spin-charge dynamics of NV$^{0}$-NV$^{-}$ composite system \citep{giri2018coupled,giri2019selective,dhomkar2016long,rao2020optimal}.

The concentration of NV$^{-}$ and NV$^{0}$ centers can vary in different
diamond samples depending on the preparation conditions and processing
parameters \citep{manson2018nv}. In fact, for a given diamond substrate
there can be a local variation of charge state ratio depending on
the local crystal strain, impurity, crush force and so on \citep{mccormick1997characterization}.
Moreover, as a function of various parameters like optical power,
magnetic field, illumination wavelength \citep{giri2018coupled,aslam2013photo},
temperature \citep{chen2011temperature} etc., inter-conversion mechanisms
of ionization and recombination \citep{aslam2013photo} are involved
which cause NV$^{-}$ $\rightarrow$ NV$^{0}$ and NV$^{0}$ $\rightarrow$
NV$^{-}$conversions, respectively. Since the charge state conversion
(CSC) influences the NV spin relaxation behaviors, and hence its sensing
efficiency, it is required to tailor the CSC dynamics by tuning the
above mentioned parameters in such a way that the sensitivity is maximized.
In this context, it is necessary to obtain a quantitative picture
of NV$^{-}$ and NV$^{0}$charge states of the NV ensemble at given
experimental conditions.

Through this paper we demonstrate two easily implementable yet powerful
method to estimate the distribution of NV$^{-}$ and NV$^{0}$ for
an ensemble system embedded in a diamond substrate. Earlier reports
have demonstrated spectral decomposition for NV charge states by modifying
the NV$^{-}$ PL signal using a microwave field which is resonant
between the $m_{S}={0}$ and $m_{S}={-1}$ states \citep{craik2020microwave}
and by studying a number of diamond samples with varying concentration
of the charge states \citep{alsid2019photoluminescence}. However,
our method relies on the external magnetic field dependent optical
properties of NV centers and a novel optical filter-based decomposition
spectroscopy protocol. The protocol has the flexibility that it can
be applied for different diamond samples and at various experimental
conditions like, temperature, magnetic field, optical power, illumination
wavelength and so on. Thus for a given NV ensemble, one can record
the change in NV$^{-}$ and NV$^{0}$ distributions as a function
of the physical conditions and determine the change in charge state
distribution.

Based on the fact that an off-axis external magnetic field reduces
the NV$^{-}$ spin polarization and hence the PL intensity of the
optical spectra, we have performed spectral decomposition to separate
out the spectrum of NV$^{0}$ and NV$^{-}$ charge states. Next, we
determine the transmissivity of the PL signal from NV$^{0}$ and NV$^{-}$
through an optical filter and use it in our spectral decomposition
protocol to create charge state mapping images for NV$^{0}$ and NV$^{-}$
centers.

\section{Experimental method}

\begin{figure}
\includegraphics[bb=118.35bp 69.0106bp 789bp 394.346bp,width=1\columnwidth]{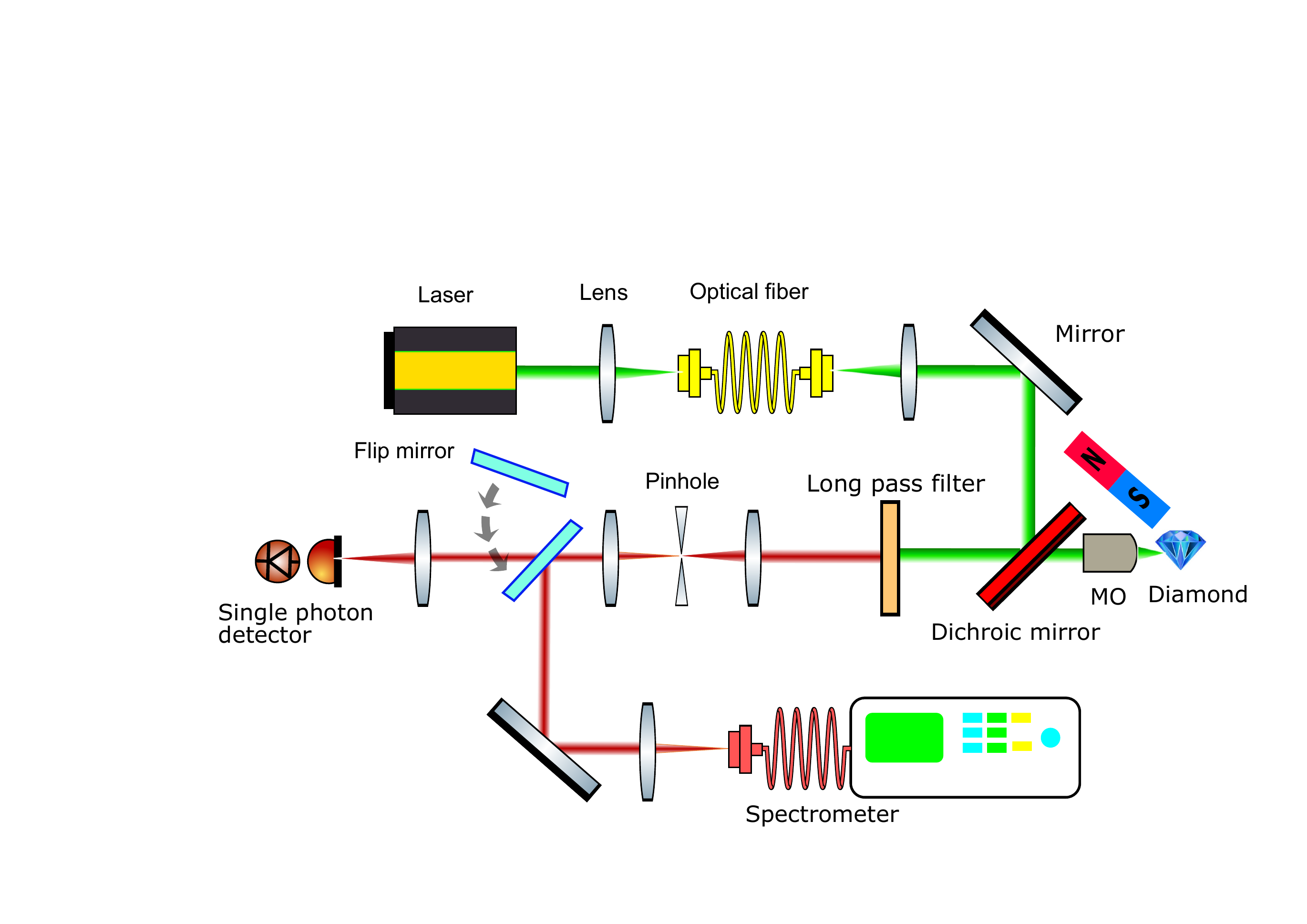}\caption{Schematic diagram of the home-built confocal set-up which is combined
with a spectrometer. A permanent magnet was used to apply the magnetic
field to the NV system}
\label{set up}
\end{figure}

Optical and magnetic resonance spectroscopy experiments were performed
in a home-built experimental setup where confocal microscopy is combined
with optical spectroscopy. We measured ensembles of NVs embedded in
a type-Ib single crystal diamond substrate. The NV concentration for
this sample is 20 ppm. Fig. \ref{set up} shows a simplified schematic
diagram of our experimental set-up. We use a diode-pumped 532 nm solid
state laser for exciting the NV centers. A high numerical aperture
(NA=1.3) microscope objective (MO) tightly focuses the excitation
beam onto the diamond sample. For creating PL mapping images of the
sample, the MO is mounted on a nano-positioning piezo stage which
has a traveling range of $100\mu m\times100\mu m$ in XY plane and
$20\mu m$ along the Z axis. The PL signal from the sample is collected
by the same MO, is separated from the reflected laser signal by a
dichroic mirror and a long pass 550 nm filter and passes through a
pinhole which rejects any out-of-focus signal. Next the signal is
collimated using a lens and detected by a Si avalanche photodiode
(APD) which is sensitive at the single photon level. For optical spectroscopy
measurements, we use a flip mirror to couple the signal to the input
slit of a CCD spectrometer through an optical fiber.

\section{Decomposition of optical spectra into NV$^{0}$ and NV$^{-}$ }

The PL emission from an NV center varies with the strength and orientation
of an external magnetic field \cite{lai2009influence,tetienne2012magnetic,epstein2005anisotropic}.
The dependence differs for the two relevant charge states, as discussed
below. In the absence of a magnetic field, the NV$^{-}$ system is
initialized to the $m_{S}={0}$ state by optical illumination. However,
application of a magnetic field that is not parallel to the NV symmetry
axis influences the optical pumping process which results in a redistribution
of population in the electronic spin levels $m_{S}={0}$ and $m_{S}={\pm1}$
which is termed as spin mixing \citep{lai2009influence,tetienne2012magnetic}.
An increase in the strength of the applied field enhances the spin
mixing. Hence, the probability of non-radiative transitions from the
excited $m_{S}={\pm1}$ states to the ground state through the metastable
singlet state increases, which leads to a decrease in the NV$^{-}$
PL signal \citep{tetienne2012magnetic,lai2009influence}. Application
of a 600 G field along the {[}100{]} axis of a diamond crystal completely
depolarizes the electronic spins which can be referred to a fully
spin mixed state. However, a change in the applied magnetic field
does not have an impact on the the ionization-recombination dynamics
\citep{giri2019selective}. Thus the distribution of population between
the NV$^{0}$ and NV$^{-}$ charge states is not influenced by the
applied field \citep{giri2019selective,capelli2017magnetic}. It has
been observed that the change in strength of the magnetic field does
not affect the NV$^{0}$ PL intensity for an NV ensemble \cite{giri2019selective}.

We have used this charge state selective change in PL signal due to
an off axis magnetic field to separate the contributions from NV$^{0}$
and NV$^{-}$ to the measured PL spectra for an NV ensemble. Under
constant excitation with a 532 nm laser, we have measured spectra
of an NV ensemble at different strengths of the magnetic field that
is not paralleled to the NV axis. We measure NV centers whose axes
are equally distributed along the four possible crystallographic directions
in a diamond crystal oriented in {[}100{]} direction. To reduce the
effect of optical and electrical noise, each measurement is averaged
over 3000 nominally identical spectra. For each scan, the measurement
time is 10 ms and the laser intensity is 4.8 mW$\mu$m$^{-2}$. To
make sure that the applied field results in notable spin mixing and
we can capture an observable change in NV$^{-}$ PL as a function
of the field, we varied the magnetic field through a large range ($\approx$
800 G) in steps, keeping its direction fixed, and measured the spectra.
The spectra captured at field values that give the largest change
in NV$^{-}$ PL and therefore the best signal-to-noise ratio (SNR),
are used in the analysis. We denote the spectra measured at the low
and high field as $A_{low\textbf{B}}(\lambda)$ and $A_{high\textbf{B}}(\lambda)$
respectively; they are shown in Fig. \ref{fig:Spectra}(a).

We write the spectrum at low field ($\approx$ 170 G) as a composition
of the NV$^{-}$ and NV$^{0}$ spectra $A_{NV^{0}}(\lambda)$ and
$A_{NV^{-}}(\lambda)$ respectively, in Eq. \ref{eq:1}.

\begin{eqnarray}
A_{low\textbf{B}}(\lambda) & = & A_{NV^{0}}(\lambda)+A_{NV^{-}}^{low\textbf{B}}(\lambda)\label{eq:1}
\end{eqnarray}

We aim to decompose $A_{low\textbf{B}}(\lambda)$ into $A_{NV^{0}}(\lambda)$
and $A_{NV^{-}}^{low\textbf{B}}(\lambda)$. In $A_{high\textbf{B}}(\lambda)$,
the NV$^{-}$ PL intensity is selectively modified by enhancing spin
mixing with a higher magnetic field, while the NV$^{0}$ signal is
not significantly affected, as discussed earlier. Hence, one can express

\begin{eqnarray}
A_{high\textbf{B}}(\lambda) & = & A_{NV^{0}}(\lambda)+A_{NV^{-}}^{high\textbf{B}}(\lambda),\label{eq:2}
\end{eqnarray}

where $A_{NV^{-}}^{high\textbf{B}}(\lambda)$ is the PL contribution
to the total spectrum from NV$^{-}$ centers at high field. We quantify
the change in the NV$^{-}$ signal as $NV_{diff}^{-}$ as a result
of changing the field by subtracting $A_{high\textbf{B}}(\lambda)$
from $A_{low\textbf{B}}(\lambda$ ):

\begin{eqnarray}
\begin{split}NV_{diff}^{-}=A_{low\textbf{B}}(\lambda)-A_{high\textbf{B}}(\lambda)\\
=A_{NV^{-}}^{low\textbf{B}}(\lambda)-A_{NV^{-}}^{high\textbf{B}}(\lambda)\label{eq:3}
\end{split}
\end{eqnarray}
.

The inset Fig.\ref{fig:Spectra}(a) exhibits that $NV_{diff}^{-}$
does not show any signature of NV$^{0}$ ZPL. The inset of Fig. \ref{fig:Spectra}(a)
shows an enlarged view of the ZPL for $A_{low\textbf{B}}(\lambda)$,
$A_{high\textbf{B}}(\lambda)$ and $NV_{diff}^{-}$ and in presence
of the low and high magnetic field, where one can observe hardly any
difference in ZPL intensity between these two cases. Since the $NV_{diff}^{-}$
data contains PL contribution solely from NV$^{-}$ centers, it is
possible to determine a scaling factor $f$, a real positive number,
such that the Eq. \ref{eq:4} and \ref{eq:5} are satisfied. There
is no evidence of change in the shape of NV spectra as a function
of the strength of the applied field. We therefore assume that the
factor $f$ does not depend on $\lambda$.
\begin{flushleft}
\begin{eqnarray}
A_{NV^{-}}^{low\textbf{B}}(\lambda) & = & f*NV_{diff}^{-}\label{eq:4}\\
A_{NV^{0}}(\lambda) & = & A_{low\textbf{B}}(\lambda)-f*NV_{diff}^{-}\label{eq:5}
\end{eqnarray}
 
\begin{figure}
\includegraphics[width=1\columnwidth]{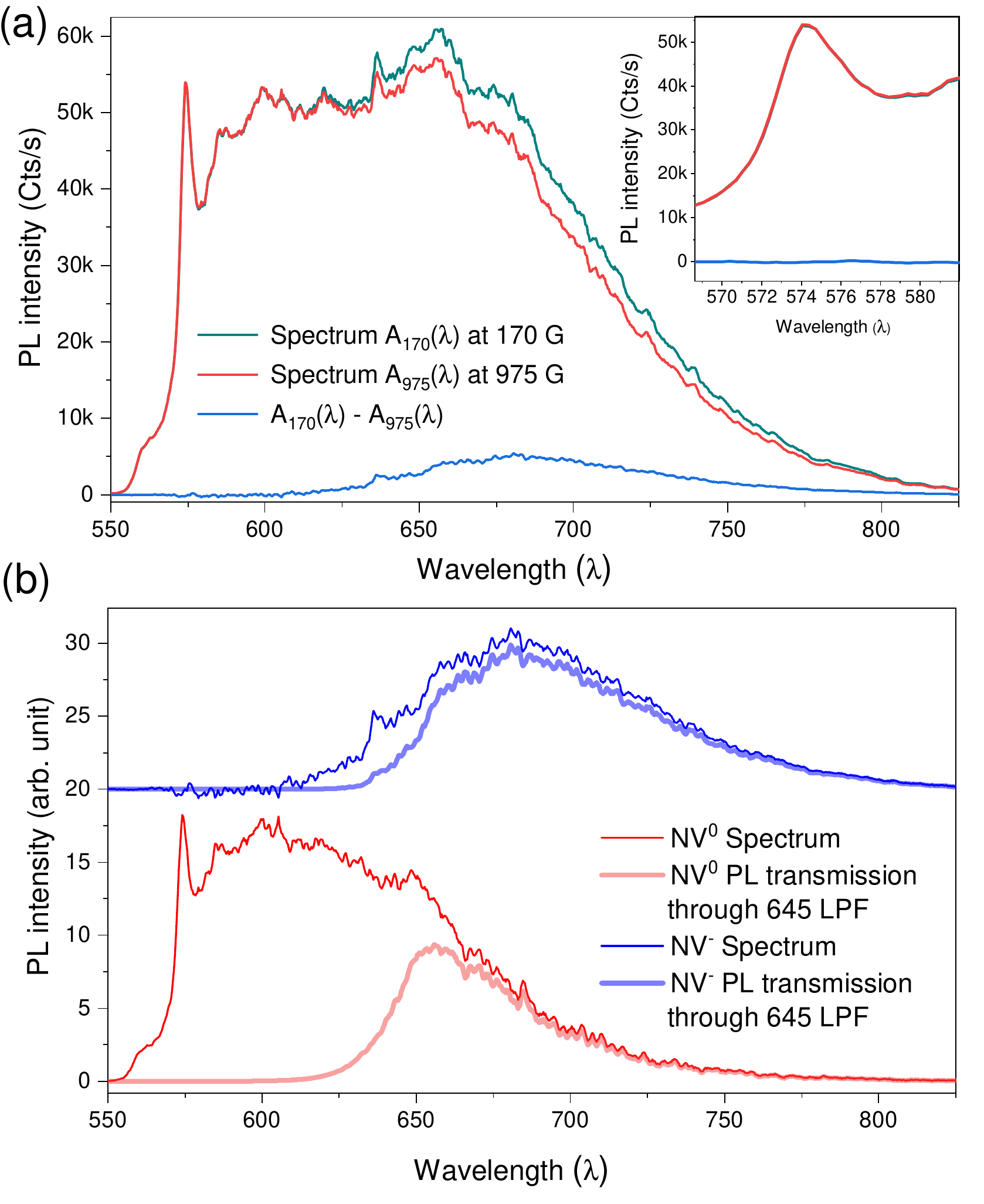}\caption{(a) Measured NV emission spectra at low ($\sim$ 170 G) and high ($\sim$
975 G) magnetic field shown by the green and red curves respectively.
The spectra show the ZPL for NV$^{0}$ at 575 nm and for NV$^{-}$
at 637 nm. The blue curve denotes the spectrum that is obtained by
subtracting the spectrum at high field from the one recorded at low
field. The inset shows an expanded view of the region near the NV$^{0}$
ZPL. (b) Extracted NV$^{0}$ spectrum (thin red curve), its transmission
through a 645 nm long pass filter (thick red curve), extracted NV$^{-}$
spectrum (thin blue curve) and its transmission through a 645 nm long
pass filter (thick blue curve). The NV$^{-}$ spectrum and its transmission
through the filter are manually shifted along the y-axis to make all
the four spectra visible in (b). \label{fig:Spectra}}
\end{figure}
\par\end{flushleft}

Proper evaluation of the factor $f$ determines the efficiency of
the decomposition analysis by assuring that $A_{NV^{0}}(\lambda)$
and $A_{NV^{-}}^{low\textbf{B}}(\lambda)$ do not have any contribution
from NV$^{-}$ and NV$^{0}$ PL, respectively. For this purpose, we
performed the operations described in Eq. \ref{eq:4} and \ref{eq:5}
for a range of values of $f$ and found $f=6.2$ to be optimal. A
higher or lower values result in a dip or peak at 637 nm, the ZPL
of NV$^{-}$, in the $A_{NV^{0}}(\lambda)$ curve. Using $f=6.2$
in Eq. \ref{eq:4} and \ref{eq:5} we can obtain the individual spectra
$A_{NV^{-}}^{low\textbf{B}}(\lambda)$ and $A_{NV^{0}}(\lambda)$,
which are shown in Fig. \ref{fig:Spectra}(b).

We have performed the decomposition of NV spectra into NV$^{0}$ and
NV$^{-}$ for a range of magnetic field values and have used these
results to study the field dependence of the PL from NV$^{0}$ and
NV$^{-}$. This analysis also provides us a quantitative estimation
of the scaling factor $f$ for a change in magnetic field from $B_{1}$
to $B_{2}$, where $B_{2}>B_{1}$. We express the NV spectra $S_{NV}(\lambda)$
as a linear combination of the contributions from NV$^{0}$ and NV$^{-}$
emission in the following way:

\begin{eqnarray}
S_{NV}(\lambda;B) & = & C_{0}(B)\hat{S}_{NV^{0}}(\lambda)+C_{-}(B)\hat{S}_{NV^{-}}(\lambda),\label{eq:8}
\end{eqnarray}

$\hat{S}_{NV^{0}}(\lambda)$ and $\hat{S}_{NV^{-}}(\lambda)$ are
the normalized base spectra, that are derived from the NV$^{0}$ and
NV$^{-}$ spectra we obtained earlier in this section by analyzing
the spectroscopic results at magnetic fields of 170 G and 975 G, such
that $\int\hat{S}_{NV^{0}}(\lambda)d\lambda=\int\hat{S}_{NV^{-}}(\lambda)d\lambda=1$.
$C_{0}(B)$ and $C_{-}(B)$ quantify the contributions to the PL spectra
from NV$^{0}$ and NV$^{-}$ states.

By using Eq. \ref{eq:4}, \ref{eq:5} and \ref{eq:8} it is straightforward
to express $f$ for a change in magnetic field from $B_{1}$ to $B_{2}$
as

\begin{eqnarray}
f(B_{1};B_{2}) & = & \frac{C_{-}(B_{1})}{C_{0}(B_{1})+C_{-}(B_{1})-C_{0}(B_{2})-C_{-}(B_{2})}.\label{eq:9}
\end{eqnarray}

\begin{figure}
\includegraphics[width=1\columnwidth]{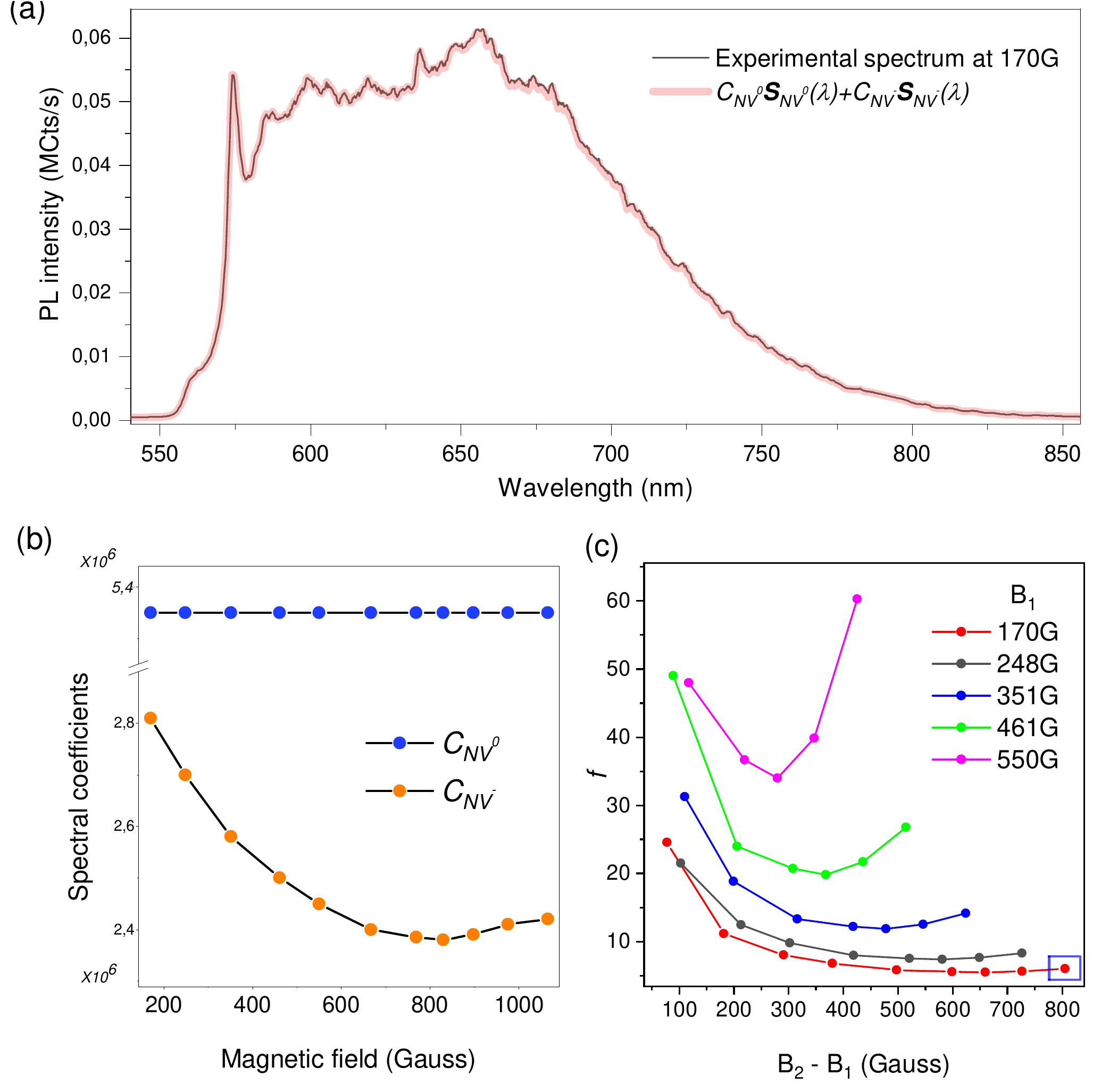}\caption{(a) The NV spectra (red curve) measured at an applied field of 170G
and its fit to $S_{NV}(\lambda;B)$ (black curve). (b) Variation of
NV$^{-}$ and NV$^{0}$ PL as a function of applied magnetic field
strength. (c) Dependence of $f(B_{1};B_{2})$ on magnetic field variation.}
\label{NVCoefficients}
\end{figure}

We determined $C_{0}(B)$ and $C_{-}(B)$ by fitting the PL spectra
measured at different magnetic fields to $S_{NV}(\lambda;B)$. The
experimental spectra measured at 170 G and its fit to $S_{NV}(\lambda;B)$
is shown in Fig.\ref{NVCoefficients}(a). We fit $S_{NV}(\lambda;B)$
to the spectra measured at different magnetic fields and have estimated
the values of $C_{0}(B)$ and $C_{-}(B)$ which are shown in Fig.\ref{NVCoefficients}(b).
The plot suggests that NV$^{-}$ PL decreases with increasing magnetic
field strength which supports spin-mixing induced PL reduction discussed
earlier in this section. At a certain field value, the NV$^{-}$ PL
reaches a minimum which indicates the fully spin-mixed state. However
with further increase in magnetic field, we see a slight enhancement
in NV$^{-}$ PL which can happen due to the influence of magnetic
field on the singlet to triplet state transition rate at the excited
state of the NV \cite{capelli2017magnetic}. It is worth noting that
there is not any observable change in the PL from NV$^{0}$ as a function
of magnetic field, which allows us to rewrite Eq. \eqref{eq:9} as

\begin{eqnarray}
f(B_{1};B_{2}) & = & \frac{C_{-}(B_{1})}{C_{-}(B_{1})-C_{-}(B_{2})}.\label{eq:10}
\end{eqnarray}
 To quantify the factor $f(B_{1};B_{2})$ at different magnetic field
regimes and intervals, we use the obtained values of $C_{0}(B)$ and
$C_{-}(B)$ in Eq. \ref{eq:10} and have calculated $f(B_{1};B_{2})$.
$B_{1}$ is varied from 170G to 550G, where $B_{2}$ is varied from
248G to 975G. The complete spin mixing, i.e. total depolarization
of the electronic spins, happens at 829 G which corresponds to the
minima in the field dependent $C_{NV^{-}}$ values shown in Fig. \ref{NVCoefficients}(b)
and $f$ values shown in Fig. \ref{NVCoefficients}(c). One can see
from Fig. \ref{NVCoefficients}(c), that $f(B_{1};B_{2})$ decreases
with increasing field difference until the fully spin mixed state
is reached. Afterwards, $f(B_{1};B_{2})$ increases with increasing
$(B_{2}-B_{1})$ due to the NV excited state singlet-triplet transition
rate dependence on magnetic field \cite{capelli2017magnetic}. Note
that for the same variation in magnetic field, $f(B_{1};B_{2})$ can
take different values depending on $B_{1}$. The data point representing
the value of $f(170G;975G)\thickapprox6.2$, which we used in our
analysis, is denoted by a square box. A lower value of $f(B_{1};B_{2})$
signifies a considerable variation in NV$^{-}$ PL upon changing the
magnetic field, which suggests that the spectra measured at 170G and
975G provided a good SNR in our NV$^{-}$- NV$^{0}$ spectral decomposition
analysis.

\begin{figure}
\includegraphics[width=1\columnwidth]{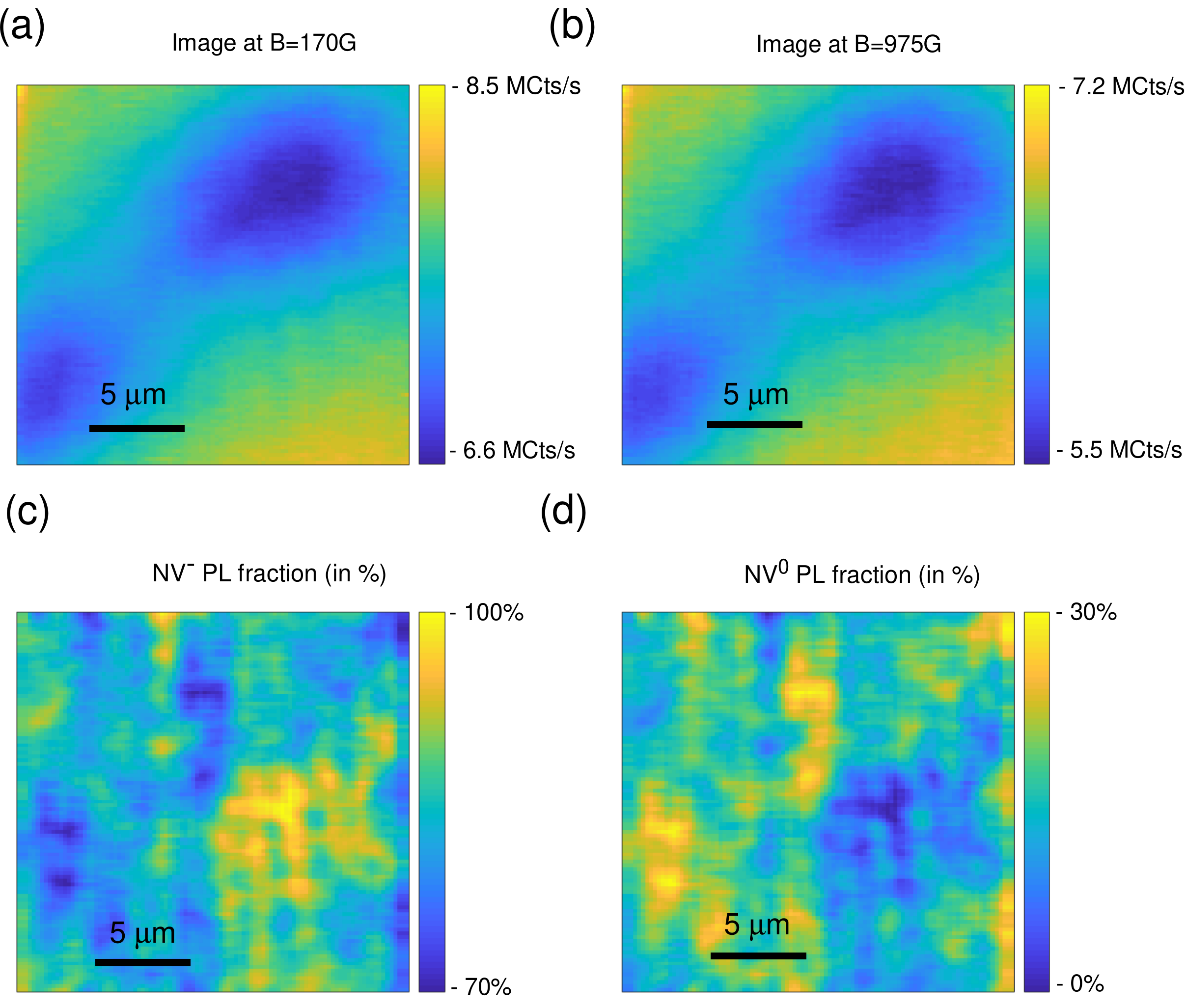} \caption{PL maps for an NV ensemble at (a) 170 G and (b) 975 G magnetic fields.
Contribution of (c) NV$^{0}$ and (d) NV$^{-}$ PL to the map shown
in (a).}
\label{fig:Image_B}
\end{figure}

The above protocol of separating out the NV$^{-}$ and NV$^{0}$ spectral
components can be used to perform charge state imaging for NV centers.
We assume $I^{lowB}(X,Y)$ and $I^{highB}(X,Y)$ as shown in Fig.\ref{fig:Image_B}(a)
and (b) are spatial maps captured for an NV ensemble in low(170G)
and high(975G) magnetic field and we aim to decompose $I^{lowB}(X,Y)$
into the maps for NV$^{-}$ and NV$^{0}$. To enhance the SNR, at
each field the maps were collected four times and added up. Provided
that apart from NV centers, there are no other defect centers or other
impurities, which generate PL under 532 nm laser excitation, one can
write

\begin{eqnarray}
I^{lowB}(X,Y) & = & I^{NV^{0}}(X,Y)+I^{NV^{-}}(X,Y),\label{eq:6}
\end{eqnarray}

where $I^{NV^{0}}(X,Y)$ and $I^{NV^{-}}(X,Y)$ are the PL matrices
that consist of contribution of PL from NV$^{0}$ and NV$^{-}$ centers
only. Following the arguments made earlier in this section, we can
calculate $I^{diff}(X,Y)$, as

\begin{eqnarray}
I^{diff}(X,Y) & = & I^{lowB}(X,Y)-I^{highB}(X,Y),\label{eq:7}
\end{eqnarray}
which should be a pure NV$^{-}$ PL signal.

Assuming that the value $f=6.2$ does not vary over the region of
microscopy, we have created maps of the charge states NV$^{-}$ and
NV$^{0}$, $I^{NV^{-}}(X,Y)=f\times{I^{diff}(X,Y)}$ and $I^{NV^{0}}(X,Y)=I^{lowB}(X,Y)-f\times{I^{diff}(X,Y)}$
and have shown their contribution to $I^{lowB}(X,Y)$ in Fig.\ref{fig:Image_B}(c)
and (d).

\section{Optical-filter based spectral decomposition}

\begin{figure}
\includegraphics[width=1\columnwidth]{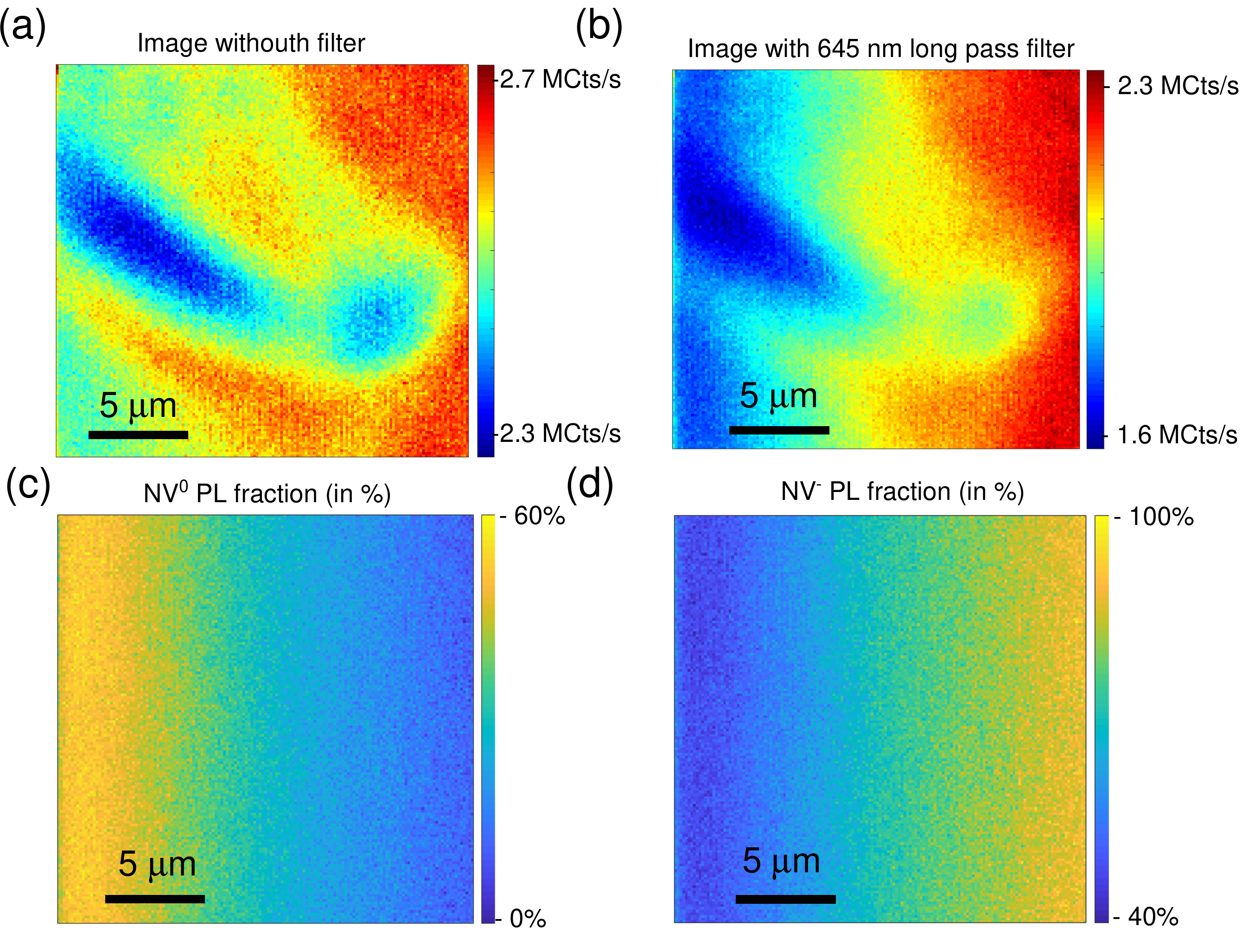}\caption{Measured PL mapping images (a) without and (b) with the 645 nm long
pass filter. Mapping image of the fraction of PL signal contribution
from (c) NV$^{0}$ (d) NV$^{-}$ centers into the image shown in (a),
calculated using our decomposition method. \label{fig:images}}
\end{figure}

Here we demonstrate a protocol to decompose the steady state NV$^{0}$
and NV$^{-}$ signals and construct separate PL images for these two
charge states. We captured the PL mapping image of an ensemble of
NV centers extended over an area of 20 $\mu m$ $\times$ 20$\mu m$
which we denote as $M_{0}(X,Y)$ and show in Fig. \ref{fig:images}(a).
We captured spectra at different spatial positions of this NV ensemble
and observed the clear signature of the ZPLs for NV$^{-}$ and NV$^{0}$.
Hence, the 20$\mu m$ $\times$ 20$\mu m$ PL mapping image consists
of signal contributions both from NV$^{-}$ and NV$^{0}$, which we
like to separate using our method. We express

\begin{eqnarray}
M_{0}(X,Y) & = & NV_{map}^{0}(X,Y)+NV_{map}^{-}(X,Y),\label{eq:11}
\end{eqnarray}
which signifies that $M_{0}$ is the sum of the two component signals
$NV_{map}^{0}(X,Y)$ and $NV_{map}^{-}(X,Y)$, which map the signal
from NV$^{0}$ and NV$^{-}$ centers, respectively. Next, we include
a 645 $nm$ long pass filter (LPF), which has different transmissions
for the signal from NV$^{0}$ and NV$^{-}$ and record a PL image
of the same 20$\mu m$ $\times$ 20$\mu m$ area. We call this image
matrix $M_{LPF}(X,Y)$ and show it in Fig.\ref{fig:images}(b). The
645 LPF transmission function for a range of $\lambda$ between 550
$nm$ and 850 $nm$ is

\begin{eqnarray}
F_{LPF}(\lambda) & = & 0.9/(1+e^{-(\lambda-645)/6.9})\label{eq:12}
\end{eqnarray}

We write $t^{0}$ and $t^{-}$ for the average transmissivity of the
PL signal from NV$^{0}$ and NV$^{-}$ through the LPF. We then express
$M_{LPF}(X,Y)$ as

\begin{eqnarray}
M_{LPF}(X,Y) & = & t^{0}NV_{map}^{0}(X,Y)+t^{-}NV_{map}^{-}(X,Y)\label{eq:13}
\end{eqnarray}

To calculate $t^{0}$ and $t^{-}$, we use the decomposed spectra
of NV$^{0}$ and NV$^{-}$ shown by the thin red and blue curves in
Fig. \ref{fig:Spectra}(b) and the thick red and blue curves $A_{NV^{0}}^{LPF}(\lambda)$
and $A_{NV^{-}}^{LPF}(\lambda)$ in Fig. \ref{fig:Spectra}(b), which
represent the spectra after the 645 LPF. Thus, the ratio of the integrated
area under the filter modulated curve to the corresponding original
spectrum gives us the filter transmitivity:

\begin{eqnarray}
t^{0} & = & \int_{550}^{850}A_{NV^{0}}^{LPF}(\lambda)d\lambda/\int_{550}^{850}A_{NV^{0}}(\lambda)d\lambda\label{eq:14}
\end{eqnarray}

\begin{eqnarray}
t^{-} & = & \int_{550}^{850}A_{NV^{-}}^{LPF}(\lambda)d\lambda/\int_{550}^{850}A_{NV^{-}}^{low\textbf{B}}(\lambda)d\lambda\label{eq:15}
\end{eqnarray}

Upon solving Eq.\ref{eq:11} and \ref{eq:13} it is straightforward
to calculate $NV_{map}^{0}(X,Y)$ and $NV_{map}^{-}(X,Y)$ which are
given by

\begin{eqnarray}
NV_{map}^{0}(X,Y) & = & \frac{M_{LPF}(X,Y)-t^{-}M_{0}(X,Y)}{t^{0}-t^{-}}\label{eq:16}
\end{eqnarray}

\begin{eqnarray}
NV_{map}^{-}(X,Y) & = & \frac{t^{0}M_{0}(X,Y)-M_{LPF}(X,Y)}{t^{0}-t^{-}}\label{eq:17}
\end{eqnarray}

We perform the integration mentioned in Eq.\ref{eq:14} and \ref{eq:15}
in the wavelength range of 550 to 850 $nm$, which covers $>99\%$
of the emission spectra (The CCD arrays in the spectrometer covers
the wavelength range of 200-1160 nm.). The resulting values are $t^{0}=0.3$
and $t^{-}=0.8$. Using these values, we obtained the separate maps
$NV_{map}^{0}(X,Y)$ and $NV_{map}^{-}(X,Y)$. Fig.\ref{fig:images}(c)
and \ref{fig:images}(d) show the relative contributions of $NV_{map}^{0}(X,Y)$
and $NV_{map}^{-}(X,Y)$ to $M_{0}(X,Y)$.

\begin{figure}
\includegraphics[width=1\columnwidth]{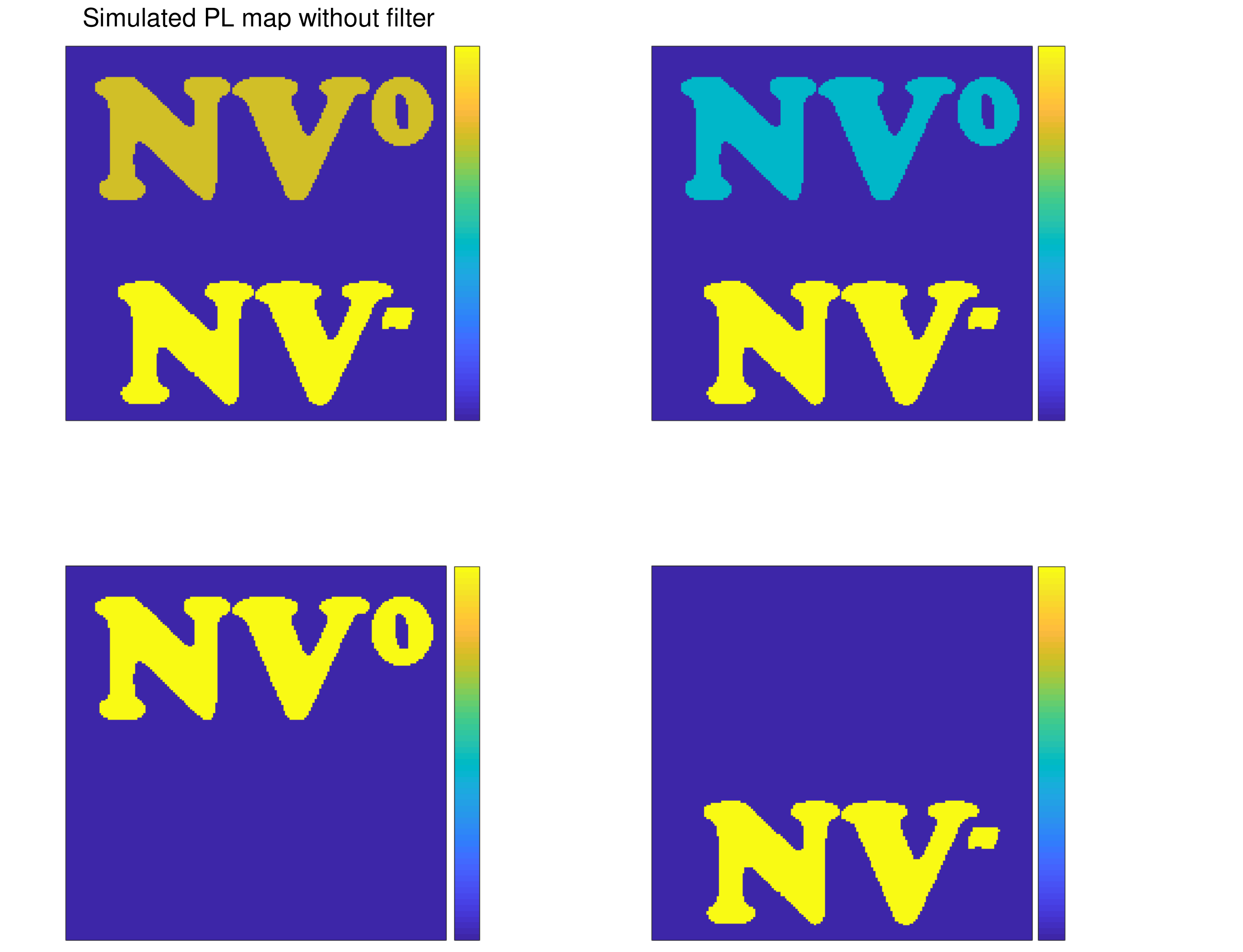}\caption{Simulated PL mapping images (a) without and (b) with the 645 nm long
pass filter. The mapping images for (c) NV$^{0}$ and (d) NV$^{-}$
centers decomposed using Eq.\ref{eq:16} and \ref{eq:17} . \label{fig:simulated-PL-mapping}}
\end{figure}

As an additional check of this method, we decompose a simulated PL
matrix which contains signal both from NV$^{0}$ and NV$^{-}$. Fig.\ref{fig:simulated-PL-mapping}(a)
represents the simulated PL mapping image $M_{0}(X,Y)$, where the
pixels inside the letters 'NV$^{0}$', 'NV$^{-}$' are assigned with
PL signals from NV$^{0}$($PL_{NV^{0}}$), NV$^{-}$($PL_{NV^{-}}$)
and rest of the pixels have the value zero. These values of the PL
signal represent total count rates. Now we simulate a PL mapping image
$M_{LPF}(X,Y)$ assuming that we place a 645 LPF before the detector,
while keeping other conditions unchanged. This is expressed by a PL
matrix where the pixels are $t^{0}*PL_{NV^{0}}$ and $t^{-}*PL_{NV^{-}}$.
From Eq.\ref{eq:16} and \ref{eq:17}, using the values of $t^{0}$
and $t^{-}$, we obtain the matrices $NV_{map}^{0}(X,Y)$ and $NV_{map}^{-}(X,Y)$.
Fig.\ref{fig:simulated-PL-mapping}(c) and \ref{fig:simulated-PL-mapping}(d)
show these maps as fractional contributions to $M_{0}(X,Y)$. One
can observe that Fig.\ref{fig:simulated-PL-mapping}(c) comprises
NV$^{0}$ PL of equal intensity at the pixel positions at 'NV$^{0}$'
like in $M_{0}(X,Y)$, whereas Fig.\ref{fig:simulated-PL-mapping}(d)
contains only NV$^{-}$ PL at the corresponding pixels as $M_{0}(X,Y)$.
The pixels with zero value are zero in both $NV_{map}^{0}(X,Y)$ and
$NV_{map}^{-}(X,Y)$. Hence, we conclude that our filter-based decomposition
protocol can perform charge state imaging for NV centers with full
efficiency. However, the simulated image $M_{0}(X,Y)$ contained fully
spatially resolved signal from NV$^{0}$ and NV$^{-}$ which allowed
us spatial decomposition of the charge states. For this test, we have
assumed that there are no other sources of PL under 532 nm excitation
apart from NV centers in one of the 2 charge states that contribute
to the PL. However, in case of other PL sources, like other color
centers, it is possible to modify the method by using additional optical
filters. Thus the method has the flexibility to decompose a PL map
which is composed of spectrum of different origins and create PL mapping
images for the concerned elementary color centers.

\section{Conclusion}

In conclusion, through this paper we have demonstrated two novel techniques:
a spectral decomposition technique to decompose the spectrum of NV$^{-}$
and NV$^{0}$, and a deconvolution protocol to create separate PL
imaging for NV$^{-}$ and NV$^{0}$. Our spectral decomposition technique
is based on the response of the PL signal from NV$^{-}$ charge state
under the influence of an off-axis magnetic field, whereas NV$^{0}$
center does not show any observable change in PL as a function of
magnetic field. Importantly, for NV spectra measured for different
diamond samples and different magnetic fields, our spectral decomposition
method remains applicable. Subsequently, relying on the fact that
a 645 nm long pass filter modulates the NV$^{-}$ and NV$^{0}$ signal
in different ways, we have applied a spectral decomposition protocol
and created PL mapping images for NV$^{-}$ and NV$^{0}$ centers
separately. The method turns out to be fully efficient to decompose
when we test it on a simulated mapping image having signal both from
NV$^{-}$ and NV$^{0}$.

\bibliographystyle{apsrev4-1}
\bibliography{draft}

\end{document}